\documentclass[12pt]{article}
  \usepackage{pstricks,pstricks-add,pst-math,pst-xkey}
  \usepackage[english]{babel}
  \usepackage{algpseudocode,algorithmicx, algorithm}
  \usepackage[utf8x]{inputenc}
  \usepackage{amsmath,amsfonts,amsthm, amssymb,ae, enumerate}
  \usepackage{dsfont}
  \usepackage{graphicx}
  \usepackage{boxedminipage}
  \usepackage{epstopdf}
  \usepackage{setspace}
  \usepackage{bm} 					
  \usepackage{natbib}
  \usepackage{caption, subcaption}			
  \usepackage{mdframed}				
 \usepackage{multirow}	
 \usepackage{rotating}
 \usepackage{comment}

  \stepcounter{page}
  \newtheorem{theorem}{Theorem}[section]
  \newtheorem{lemma}[theorem]{Lemma}

  \def\ds{\displaystyle}

\begin{document}

\title{\bf Flexible Bayesian modelling in dichotomous item response theory using mixtures of skewed item curves}
\author{F. B. Gon\c{c}alves, J. Venturelli S. L., R. H. Loschi}
\date{}

\maketitle

\begin{center}
{\footnotesize Universidade Federal de Minas Gerais, Brazil}
\end{center}

\renewcommand{\abstractname}{Abstract}
\begin{abstract}

Most Item Response Theory (IRT) models for dichotomous responses are based on probit or logit link functions which  assume a symmetric relationship between the probability of a correct response and the latent traits of individuals submitted to a test. This assumption restricts the use of those models to the case in which all items have a symmetric behaviour. On the other hand, asymmetric models proposed in the literature impose that all the items in a test have an asymmetric behaviour. This assumption is inappropriate for great part of the tests which are, in general, composed by both symmetric and asymmetric items. Furthermore, a straightforward extension of the existing models in the literature would require a prior selection of the items' symmetry/asymmetry status.
This paper proposes a Bayesian IRT model that accounts for symmetric and asymmetric items in a flexible though parsimonious way. That is achieved by assigning a finite mixture prior to the skewness parameter, with one of the mixture components being a point-mass at zero. This allows
for analyses under both model selection and model averaging approaches. Asymmetric item curves are designed through the centred skew normal distribution, which has a particularly appealing parametrisation in terms of parameter interpretation and computational efficiency. An efficient MCMC algorithm is proposed to perform Bayesian inference and its performance is investigated in some simulated examples. Finally, the proposed methodology is applied to a data set from a large scale educational exam in Brazil.


\end{abstract}

\section{Introduction}

The increasing competitiveness for places at the best universities and the growing use of assessments through computer-based tests are some examples that motivate the  building of  educational tests with more challenging items able to select individuals with the capacity of solving complex items. In psychometry, an item that requires the use of a large number of conjunctively interacting subprocesses is a complex item. Such complex items allow us to evaluate if a student has developed the ability to use the individual skills acquired in some different subject areas in a joint way to solve a complex problem.
	
To select candidates with this profile,   tests should be elaborated including not only symmetric items. Symmetric items discriminate individuals with low and high trait levels in the same way since under a symmetric item characteristic curve (ICC), the probability of a correct answer approaches zero at the same rate as it approaches one. Thus a test composed only of symmetric items will hinder an appropriate selection of outstanding individuals. To be more effective, the test should also include asymmetric items. Asymmetric items  play an important role in item response theory (IRT). Besides facilitating the identification of outstanding individuals, the asymmetry of an item works as an indicator of the degree of the underlying complexity in the item \citep{LeBo17, Boetal18} thus allowing for the evaluation of the test quality.

Item Response Theory (IRT) is a psychometric theory commonly used in educational assessments and cognitive psychology.
IRT models relate the probability of a given response in a test to the latent trait(s) of the respondent and to characteristics of the items, such as difficulty and discrimination. In an unidimensional setup, only one latent trait is considered and the ICC returns the probability of a particular response to a given item as a function of the latent trait. For more details about IRT, see \cite{rasch1960probabilistic}, \cite{birnbaum1968some}, \cite{lord1980applications}, \cite{fox2010bayesian}, \cite{embretson2013item}, \cite{bartol2015}, among others.
IRT models deal with a broad  of item formats.
We will only focus in models for dichotomous items.
The first models for dichotomous items assumed a probit \citep{albert1992bayesian, AlbertGhosh00} or a logit \citep{birnbaum1968some} function to model the ICC. This implies in a symmetric ICC.
If the goal in the test is, for example, to identify outstanding individuals with great ability to solve difficult problems, models assuming symmetric ICC are inappropriate \citep{samejima1997departure, samejima2000logistic, bazan2014extensions}.
To account for those problems, \cite{samejima1997departure, samejima2000logistic} introduces an IRT model that gives more weight
for individuals that correctly answer the most difficult items and imposes a heavier penalty to individuals that fail in the easiest items. The Logistic Positive Exponent (LPE) family of models \citep{samejima2000logistic} considers an asymmetric ICC, defined as $L(\cdot)^{\epsilon_i}$, where  $L(\cdot)$ is the logistic function, and $\epsilon_i > 0$ is the skewness parameter associated with the $i^{th}$ item. The logistic link is recovered when $\epsilon_i = 1$. Similarly, \citet{bazan2010skew} proposes the Refletion LPE (RLPE) model by defining the ICC $1-(1-L(\cdot))^{\epsilon_i}$ which, as the name says, is a reflection of the LPE curve.
A different approach is proposed in \cite{bazan2006skew}, who use the cumulative distribution function (c.d.f.) of a skew normal (SN) distribution \citep{azzalini1985class} as the ICC. This includes the symmetric ICC as a particular case when the skewness parameter is zero. This model has a similar behaviour to the LPE model when below the symmetric curve (positive asymmetry) and to the RLPE model when above the symmetric curve (negative asymmetry). See \citet{bazan2010skew} for a comprehensive understanding of practical implications of each behaviour.

Based on the model proposed by \citet{samejima2000logistic},  \citet{LeBo17} drew some important conclusions about IRT models assuming a common ICC for all individuals. Firstly,  the use of  IRT models assuming asymmetric ICC is appealing because of the growing interest in building educational tests which are able to select individuals with the ability to solve complex items. Also, the efficiency of a common metric for evaluations occurring in vertical scalings, such as across grade levels, may be compromised if item asymmetry is ignored. This possibly occurs because, in the educational systems, item complexity tends to increase by grade level. Finally, they also reinforce the findings in  \citet{Mo14} that  the misspecification of symmetric models may lead to inappropriate inference in tests that rely heavily on the item information, such as computerized adaptive testing (CAT).

The models previously mentioned are based on homogeneous tests assuming that the items are either all symmetric or all asymmetric. Usually, in a test, it is expected that a symmetric structure fits well great part of the items but not all of them. Therefore, assuming that either all items are symmetric or all items are asymmetric is bound to compromise the efficiency of the analysis. The former introduces bias to the estimates  and the latter contravenes parsimony by overfitting the data and jeopardises computational efficiency.
Moreover,  from a theoretical point-of-view, the misspecification of the link function introduces bias in the mean response estimation \citep{czadosatner92}.

As the degree of asymmetry of an item is an indicator of its complexity and negatively asymmetric items are more efficient to identify individuals with higher abilities, an ideal model should account for both item categories, symmetric and asymmetric. It should also allow us to classify each item in its appropriate category, symmetric, negatively or positively asymmetric, without penalising the computational cost.

Although the existing models may be straightforwardly adapted to set only some of the items to be asymmetric, this would require the symmetry/asymmetry status of each item to be fixed prior to the analysis and therefore, would not be of practical use. It would be ideal if IRT models could assume the asymmetric structure only for items that significantly depart from symmetry. Although, methodologies for model comparisons \citep{sahu2002bayesian} could be used to that end, a naive model selection procedure is not efficient as it would require $2^{I}$ or $3^{I}$ models to be fit and compared, where $I$ is the number of items in the test.

A robust and parsimonious way to perform model selection is to consider a model-based approach in which the full model is a finite mixture of all possible individual models. Under the Bayesian paradigm, inference is based on the posterior distribution that includes the space of all individual models in its domain and, therefore, provides the posterior probability of each model. Under several aspects, this approach is expected to be more robust and computationally more efficient than arbitrarily chosen model selection criteria \citep[see, for example,][]{gonccalves2015robust}. A well designed MCMC algorithm would automatically ignore the individual models with negligible posterior probability. The model-based Bayesian approach also allows inference to be performed under a model averaging perspective, rather than under a model selection one. This is a reasonable approach when more than one model presents significant posterior probability.

Motivated by these ideas, we introduce, in Section \ref{Se2}, an IRT model in which the ICC of each item in the test is a finite mixture of symmetric and asymmetric ICC's. The proposed finite mixture skew-probit IRT model allows to simultaneously account for symmetric and asymmetric items in the same test, as well as to identify if the items are positive or negatively asymmetric. The c.d.f. of a normal distribution is assumed to model the symmetric component of the mixture and the c.d.f. of a skew-normal distribution is adopted for the two asymmetric components. Furthermore, the skewness parameter of each of the two asymmetric curves (one positively and one negatively asymmetric) is estimated without constraints in its parametric space. Thus, the resulting models are expected to accommodate well enough all asymmetric items in a test independently of their type (positive or negative) and degree of skewness.
Compared to the existing literature on asymmetric ICC's, the main contribution of this paper lies on the fact that the proposed model is simultaneously flexible and parsimonious since it accommodates both symmetric and asymmetric ICC's in a single structure. That is attained by assuming a point-mass mixture prior distribution for the skewness parameter where a positive probability at zero accounts for the symmetric ICC and continuous distributions account for negative and positively asymmetric cases. This strategy allows us to perform model-based Bayesian model selection to conclude about the type of skewness experienced by each item. From a practical point of view, it provides an efficient way to define the complexity of the items by providing the probability of each item being symmetric, negatively or positively asymmetric.
Moreover, under a model averaging perspective, a more robust IRT model is obtained since the ICC becomes a finite mixture of the c.d.f.'s of a normal and of two skew normal distributions. Finally, the proposed model brings a considerable gain in terms of computational cost due to simplification when some items are well fitted by a symmetric ICC.

Another contribution of the paper lies on the particular parametrisation adopted for the skew-normal distributions whose c.d.f.'s are used to model the possible asymmetric behaviour of the ICC's. We use the c.d.f. of a centred skew normal (CSN) distribution \citep{azzalini1985class, AVA2008} which brings significant gain in terms of parameter interpretation and computational cost if compared to the existing alternatives - LPE, RLPE and skew-probit. A detailed explanation of this contribution is provided in Section \ref{SS21}.

The proposed mixture ICC is expected to provide a good fit to the items in most applications, whilst having the parsimony of a parametric model (each asymmetric ICC actually has only one additional parameter compared to the traditional symmetric logit/probit models). Other approaches to improve model flexibility over the traditional models is to consider non-parametric structures. Generally speaking, the comparison between parametric and non-parametric models involves trade-offs regarding features such as model flexibility, model parsimony, parameter interpretation, precision in estimation and computational complexity, and none of the two approaches is uniformly better than the other. For example, \citet{Bart2017} consider a discrete non-parametric approach to model the probability mass function (p.m.f.) of the responses. The model assumes that each individual belongs to a latent class, such that individuals in the same class have the same response probability. On one hand, the model offers flexibility by being free from the parametric structure of traditional models. On the other hand, it imposes the latent class structure to the individuals and requires complex algorithms such as reversible-jump MCMC to perform inference. Another non-parametric approach is proposed in \citet{Karaba2016}. This author considers an infinite mixture of traditional ICC models, such as the Rasch model. Whilst a mixture of curves offers considerable model flexibility, the large number of unknown parameters may compromise model parsimony.

Inference for the proposed model is performed through an efficient MCMC algorithm to sample from the joint posterior distribution of all the unknown quantities in the model. This is presented in Section \ref{Se3}. Section \ref{Se4} presents some simulated examples. An application to a large scale educational assessment exam in Brazil is presented in Section \ref{Se5} focusing on the identification of items which are more efficient in differentiating individuals with higher abilities.

\section{Proposed Model}
\label{Se2}

The proposed model assumes that the ICC of each item is a mixture of one normal and two skew normal (one positively and one negatively skewed) c.d.f.'s. We consider the centered parametrisation of the skew-normal distribution to ease parameter interpretation and to boost computational efficiency, as it is explained in Section \ref{SS21}.

\subsection{Two parameters probit centred skew normal ogive model}\label{subsmodel}

Assume that a random sample of $J$ individuals is submitted to a test composed
by $I$ items. Denote by  $\mathbf{Y} = (Y_{ij})_{I \times J}$, the $I \times J$ matrix of all responses to the test, where $Y_{ij}$ is the indicator of a correct response given to item $i$ by individual $j$. The two parameters probit centred skew normal ogive (2PCSNO) is defined as
\begin{eqnarray}\label{eq:ModeloTRI}
(Y_{ij} \rvert p_{ij}) &\overset{ind}{\sim}& Bernoulli(p_{ij}(m_{ij},\gamma_i)), \label{mmeq1}\\
p_{ij} (m_{ij},\gamma_i) &=& \Phi_{CSN}(m_{ij},\gamma_i), \label{mmeq2}
\end{eqnarray}
where $ m_{ij} = a_i(\theta_j - b_i)$ and $\Phi_{CSN}(m_{ij},\gamma_i)$ is the c.d.f. of the centred skew normal distribution, with skewness parameter $\gamma_i$, evaluated at $m_{ij}$. Moreover, $a_i$ and $b_i$ are, respectively, the discrimination and difficulty parameters of item $i$ and $\theta_j$ is the latent trait of individual $j$. Hereafter, we  consider the notation $\mathbf{\theta} = (\theta_{1},\ldots,\theta_{J})$, $\textbf{a} = (a_{1}, \ldots, a_{I})^T$, $\textbf{b} = (b_{1},\ldots,b_I)^T$, $\mathbf{\gamma} = (\gamma_1,\ldots,\gamma_I)^T$ and the analogous bold notation to define the vector of the respective components yet to be defined.

Although asymmetric ICC models are more flexible and allow for specific aims like identifying outstanding individuals, it is not reasonable to consider a test with only asymmetric items. Symmetric items are often quite efficient to estimate individuals' latent traits and, therefore, most of the items are designed to be symmetric. Given their higher statistical and computational complexity, asymmetric items should only be considered for specific aims or to provide a better fit to items that are, unexpectedly, not properly accommodated by a symmetric ICC. In any case, it is typically hard to define \emph{a priori} which items should be modelled by an asymmetric ICC. Moreover, it is not reasonable to assume that all the items in a test are asymmetric as this would typically defy the parsimony principle.

To mitigate this problem and to account for an eventual heterogeneity among the items symmetry status, we propose
a model-based approach to consider the uncertainty about each item's skewness. The finite mixture skew-probit IRT model consists of a finite mixture of ICC's in which each of the three mixture components correspond to one skewness status - symmetric, negatively and positively asymmetric. This is attained by considering the model in (\ref{mmeq1})-(\ref{mmeq2}) and  eliciting the point-mass mixture prior for the skewness parameter $\gamma_i$ hierarchically represented as

\begin{equation}\label{eq:priorGamma}
\ds
\left.\begin{array}{lll}
\gamma_i&=&\left\{
               \begin{array}{ll}
                 0, & \mbox{if } Z_i=(1,0,0) \\
                 \gamma_{i-}, & \mbox{if } Z_i=(0,1,0) \\
                 \gamma_{i+}, & \mbox{if } Z_i=(0,0,1)
               \end{array}
             \right. \;\;\;\;\;\;\;\;\;\;\;\;\textcolor[rgb]{1.00,1.00,1.00}{.}\\
		Z_i|w_i&\overset{ind}{\sim}& Multinomial(1,w_i), \;\;\;\;\;\;\;\;\;\;\;\;\textcolor[rgb]{1.00,1.00,1.00}{.} \\
		w_i&\overset{iid}{\sim}&Dirichlet(\alpha_0,\alpha_1,\alpha_2), \;\;\;\;\;\;\;\;\;\;\;\;\textcolor[rgb]{1.00,1.00,1.00}{.} \\
		-\gamma_{i-}&\overset{iid}{\sim}& Beta(\alpha,\beta), \;\;\;\;\;\;\;\;\;\;\;\;\textcolor[rgb]{1.00,1.00,1.00}{.} \\
		\gamma_{i+}&\overset{iid}{\sim}& Beta(\alpha,\beta), \;\;\;\;\;\;\;\;\;\;\;\;\textcolor[rgb]{1.00,1.00,1.00}{.}
\end{array}\right\}
\end{equation}
where the Beta distributions are truncated to the interval (0,\;0.99527) and the weights $w_i = (w_{i0},w_{i1},w_{i2})$ are such that $w_{ik} \in (0,1)$ and $\sum_{k=0}^{2}w_{ik}=1$. The point-mass mixture prior in (\ref{eq:priorGamma}) basically states that, $\gamma_i$ is zero with probability $w_{i0}$, is negative and has a continuous distribution in $(-0.99527,0)$ with probability $w_{i1}$, and is positive and has a continuous distribution in $(0,0.99527)$ with probability $w_{i2}$.

As the parameters $\gamma_{i-}$ and  $\gamma_{i+}$ are unknown, under this approach, the ICC of each of the items can be the c.d.f. of a normal distribution or the c.d.f. of a skew normal distribution with any possible level of asymmetry. This is expected to provide a good fit for all items in a test in most cases. The Beta prior distribution assumes different shapes and may accommodate different prior knowledge about the item's complexity represented by the skewness parameters.  Nevertheless, any other continuous distribution truncated to the interval (0,\;0.99527) may be used at no extra cost. Finally, the proposed hierarchical representation for the  prior of $\gamma_i$ is computationally appealing as it is shown in Sections \ref{Se3} and \ref{Se4}.

The prior in (\ref{eq:priorGamma}) can be adapted to account for situations in which the items are aggregated in homogeneous categories composed by items that share the same or similar degree of  asymmetry. To achieve this purpose, let $\gamma_{i-} = \gamma_{G-}$ and $\gamma_{i+} = \gamma_{G+}$, for all item $i$.
If there exists substantial estimation error on the item level, when applicable, this approach minimises problems related to asymmetry estimation \citep{LeBo17}. The mixture of ICC's for item $i$ can be explicitly defined by integrating out the $Z_i$'s:
\begin{equation}\label{eq:priorGamma2}
\ds P(Y_{ij}=1|w_i,\gamma_{i-},\gamma_{i+})=w_{i0}p_{ij}(m_{ij},0) + w_{i1}p_{ij}(m_{ij},\gamma_{i-}) + w_{i2}p_{ij}(m_{ij},\gamma_{i+}).
\end{equation}

The prior distribution of the weights $w_i$ play a crucial role in the 2PCSNO model. Basically, we want to be parsimonious and, therefore, only detect asymmetric items when that offers a substantially better fit than the symmetric alternative. Since the latter model is nested in the former, we need some natural way to penalise the more general model. That can be achieved in a robust way through the prior specification for the weights $w_i$ by assigning higher probabilities to the symmetric model. More specifically, we assume independent Dirichlet priors and, since each $w_i$ indexes the distribution of only one Multinomial random variable, we fix the Dirichlet hyperparameters to be smaller than 1, following the ideas from Lemma 1 in \citet{gonccalves2015robust}. In our context, such a lemma states that
\begin{lemma}\label{prop1}
For a prior distribution $w_i\sim Dirichlet(\alpha_0,\alpha_1,\alpha_2)$, the posterior mean of $w_{ik}$, for $k=0,1,2$, is restricted to the interval
$$\ds \left(\frac{\alpha_k}{1+\sum_{l=1}^3\alpha_l} , \frac{\alpha_k+1}{1+\sum_{l=1}^3\alpha_l} \right).$$
\end{lemma}
If, for example, a uniform prior ($Dirichlet(1,1,1)$) is adopted for $w_i$, the posterior mean of each model probability will be restricted to the interval $(0.25,0.5)$, compromising the model selection criterion. On the other hand, if a $Dirichlet(0.01,0.01,0.01)$ is adopted, the respective posterior means will be restricted to the interval $(0.0097,0.9805)$.
Simulated studies shown in Section \ref{Se4} indicate that a $Dirichlet(0.1;0.01;0.01)$ is a good and robust choice under different scenarios.

Prior specification is completed by assuming that $\mathbf{\theta}$, $\textbf{a}$ and $\textbf{b}$ are independent with $\theta_j \overset{iid}{\sim} N(0,1)$, $a_i \overset{iid}{\sim} N(\mu_{a_i},\sigma_{a_i}^2)\mathbb{I}(a_i > 0)$ and $b_i \overset{iid}{\sim}N(\mu_{b_i},\sigma_{b_i}^2)$.

\subsection{Extension for the three parameters model}

The 2PCSNO model can be naturally extended to include the guessing parameter. The 3PCSNO is obtained from (\ref{mmeq1})-(\ref{mmeq2}) by replacing (\ref{mmeq2}) with
\begin{equation}\label{eq:SNICC2}
\ds p_{ij}(m_{ij},\gamma_i)=c_i+(1-c_i)\Phi_{CSN}(m_{ij},\gamma_i)
\end{equation}
and $c_i\overset{iid}{\sim} Beta(\alpha_c,\beta_c)$.

From a practical point of view, though, this formulation should only be considered when large data sets are available. The known difficulties related to the parameter estimation in the three parameters classical IRT models are bound to be even more severe in the case asymmetric ICCs are considered. In particular, it may jeopardise the detection and estimation of asymmetry if not enough information is available. One practical solution is to estimate the guessing parameters under the (symmetric) two parameter normal ogive (3PNO) model and fix their values in the 3PCSNO to be the point estimates obtained under the 3PNO model. This is reasonable given the same role played by the guessing parameters in both models.

\subsection{PCSNO item characteristic curve}\label{SS21}

In order to avoid inference problems and to boost computational efficiency, the PCSNO models consider the centered parametrisation of the skew-normal distribution proposed by \citet{azzalini1985class}. We present the centred skew normal distribution and highlight the importance of considering this parametrisation to specify the ICC in the PCSNO models. In particular, we discuss its advantages when compared to the direct parametrisation considered in \cite{bazan2006skew}.

The c.d.f. $\Phi_{CSN}(u,\gamma)$ of the CSN distribution is given by
\begin{equation}\label{eq:cdfCSN}
\ds \Phi_{CSN}(u,\gamma) = \int\limits_{-\infty}^{u} \frac{2}{(1+s^2 \gamma^{2/3})} \phi\left( \frac{t - s\gamma^{1/3}}{(1+s^2 \gamma^{2/3})} \right)\Phi\left( g(\gamma)\left( \frac{t - s\gamma^{1/3}}{(1+s^2 \gamma^{2/3})}\right)\right) dt,
\end{equation}
where $r = \sqrt{2/\pi}$, $s = \left(2 /(4-\pi)\right)^{1/3}$ and $g(\gamma)  = s \gamma^{1/3}[r^2 + s^2 \gamma^{2/3}(r^2-1)]^{-1/2}$.
The parameter $\gamma$ takes value in $(-0.99527,0.99527)$ and represents the level of positive or negative asymmetry of the distribution.

The centred skew normal distribution, proposed by \citet{azzalini1985class}, has mean 0, variance 1 and better statistical properties than the non-centred skew-normal distribution whose mean and variance depend on the skewness parameter. Inference based on likelihood methods in the non-centered skew-normal family is problematic. The shape of the profile log-likelihood as a function of the skewness parameter $\lambda$ is non-quadratic and has a stationary point at $\lambda=0$ which makes maximum likelihood estimation cumbersome. Furthermore, at $\lambda=0$, the expected Fisher information is singular, even if all parameters are identifiable. This  prevents the use of maximum likelihood-based procedures for testing normality and, consequently, for ICC symmetry, under the skew-normal family. For a detailed discussion see \citet{AVA2008}.

In our context, the centred version of the skew-normal distribution has further substantial advantages when compared to the non-centred one which is used in \cite{bazan2006skew} and the LPE \citep{samejima2000logistic} and RLPE \citep{bazan2010skew} models. Under the centred parametrisation, parameters $a$ and $b$ play the same role in both the symmetric and asymmetric ICC's. Assuming the centred skew-normal distribution, the c.d.f. is given by $F(u)=\Phi_{CSN}(a(u - b),\gamma)$ and the  mean and variance are $b$ and $1/a^2$, respectively, not depending on the skewness parameter. Under the non-centred parametrisation and the LPE and RLPE models, both statistics do depend on the skewness parameters. For example, for the non-centred skew ogive model, the mean and variance, respectively, are  $\sqrt{2/\pi}(\lambda/\sqrt{1+\lambda^2})$ and $1-(2/\pi)(\lambda^2/(1+\lambda^2))$. Thus, unlike in the existing models for asymmetric ICC's, parameters $a$ and $b$ play the same role in the 2PCSNO model and, therefore, have the same interpretation. This favors parameter identification which, in turn, leads to MCMC algorithms with better convergence properties.

\begin{figure}[h!]
	\centering
	\includegraphics[width=10.5cm]{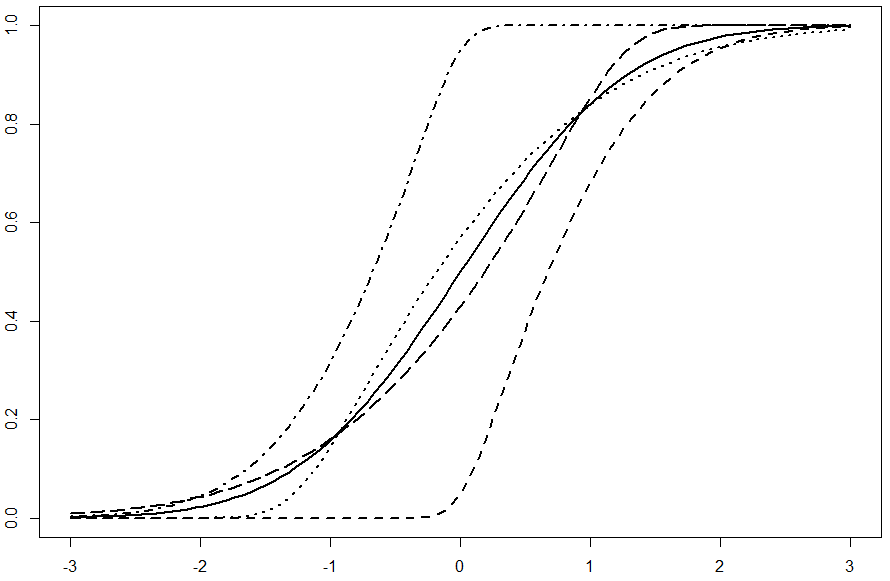}
	\caption{ICC for: Symmetric (solid), CSN $\gamma=0.9$ (dotted), SN $\gamma=0.9$ (short dashed), CSN $\gamma=-0.9$ (long dashed) and SN $\gamma=-0.9$ (dot-dashed) distributions. $a=1$ and $b=0$ for all the three curves.}
	\label{fig:comparacaoICC}
\end{figure}

Figure \ref{fig:comparacaoICC} shows the symmetric ICC, the asymmetric ICC under the centred parametrisation and the asymmetric ICC under the non-centred parametrisation.
All curves have the same discrimination and difficulty parameters and both asymmetric curves have skewness parameter set to 0.9, meaning the same level of asymmetry. Nevertheless, the asymmetric curve under the centred parametrisation is much closer to the symmetric one.

\section{Inference via MCMC}\label{Se3}

Posterior inference is discussed under the 3PCSNO model.
In order to circumvent the intractability introduced by the guessing parameter, we consider a set of auxiliary variables as proposed by \citet{bambirra2018bayesian} that preserves the original model upon marginalisation. We define the following auxiliary variables:
\begin{equation}
\ds (D_{ij}|c_i) \sim Bernoulli(c_i),\;\forall\;i,j, 
\end{equation}
and set $(Y_{ij}|D_{ij}=1)\sim Ber(1)$ and $(Y_{ij}|D_{ij}=0,\cdot)\sim Ber(\Phi_{CSN}(a_i(\theta_j-b_i),\gamma_i))$. If the 2PCSNO model is considered, the $D_{ij}$ random variables and parameters $c_i$ are all set to zero.

The main goal of the inference process is to obtain the posterior distributions of all the unknown quantities in the model, denoted  by $\mathbf{\Psi} = (\textbf{a}, \textbf{b}, \textbf{c}, \mathbf{\theta}, \mathbf{\gamma}, \mathbf{\gamma_-}, \mathbf{\gamma_+}, \textbf{Z}, \mathbf{w}, \textbf{D})$, which has joint density $\pi(\mathbf{\Psi}|\textbf{Y})$ proportional to
\begin{equation}\label{eq:Posterior}
\ds \left[\prod_{i=1}^I\left[\prod_{j=1}^J \pi(Y_{ij}|D_{ij},m_{ij},\gamma_i)\pi(D_{ij}\mid c_i)\right] \pi(Z_i\mid w_i) \pi(\gamma_i|Z_i,\gamma_{i+},\gamma_{i-})\right]\pi(\mathbf{w},\mathbf{\theta},\mathbf{\gamma_{-}},\mathbf{\gamma_{+}},\textbf{a},\textbf{b},\textbf{c}).
\end{equation}
Given the complexity and high dimensionality of this posterior distribution, we resort to an MCMC algorithm to explore this distribution via Monte Carlo.
The proposed algorithm is a Gibbs sampling with some Metropolis-Hastings (MH) steps. Aiming at having the best possible convergence properties, we choose the following blocking scheme:
\begin{equation}\label{eq:BlockScheme}
\ds \quad(\mathbf{\theta})\quad(\textbf{a},\textbf{b})\quad(\textbf{w},\textbf{Z}) \quad(\mathbf{\gamma_{-},\gamma_{+}}) \quad (\textbf{D}) \quad (\textbf{c}).
\end{equation}
Details of the proposed MCMC algorithm are presented in Appendix B.

\section{Simulated Studies}\label{Se4}

In order to evaluate the performance of the 2PCSNO we run two simulation studies. In Simulation 1, we evaluate the model sensitivity to the Dirichlet prior specification for the weights $w_i$ as well as  the impact of the number of subjects in the inference process. In Simulation 2, we compare the 2PCSNO model to the one introduced by \citet{bazan2006skew}.

\subsection{Simulation 1: Prior sensitivity analysis}

In this study, we simulate tests with 40 items which are submitted to $J=1000$, $3000$ and $10000$ subjects. We consider two scenarios. In both of them, data are generated from the model in (\ref{mmeq1})-(\ref{mmeq2}) assuming that $a_i \sim N(0,1) \mathbb{I}(a_i > 0)$ and $b_i \sim  N(0,1)$, for all $i$, and that $\theta_j \sim N(0,1)$, for all $j$. The first  scenario assumes that 27 of the items are asymmetric, with varying levels of asymmetry $\gamma_i$. In the second one, all items are symmetric, i.e., $\gamma_i=0$, for all $i$.

We adopt the following prior distributions: $a_i \sim N(1,4)$, $b_i\sim N(0,25)$, $\theta_j \sim N(0,1)$ and $-\gamma_{i-} \overset{D}{=}\gamma_{i+} \sim Beta(1,1)$ truncated on the interval (0,0.99527).

\begin{figure}[h!]
	\centering
	\includegraphics[width=0.65\textwidth]{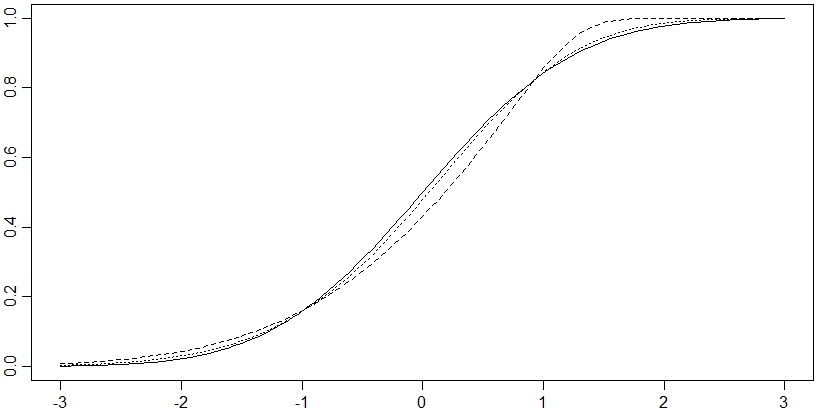}
	\caption{Symmetric (solid) and asymmetric ICC's with $\gamma=-0.4$ (dotted) and $\gamma=-0.9$ (dashed).}\label{figasm}
\end{figure}

We consider three different Dirichlet priors for the weights $w_i$. We set ${\mathbf{\alpha}}=(\alpha_0, \alpha_1, \alpha_2)=$ $ (0.01, 0.01, 0.01)$ (Pr1), $(0.05, 0.01, 0.01)$  (Pr2) and $(0.1, 0.01, 0.01)$  (Pr3). Basically, we want a prior distribution for the weights that is robust, under different characteristics of the data set, to only detect significant levels of asymmetry. As a reference, Figure \ref{figasm} shows that values of $|\gamma|$ smaller than 0.4 are considered a insignificant level of asymmetry and a clear significant level is observed for $|\gamma|>0.9$.


Results for the first scenario show that the 2PCSNO model is more efficient, in the sense of classifying item with negligible levels of asymmetry as symmetric ones, when asymmetric priors (Pr2 and Pr3) for the weights are considered. As it has been previously discussed, this contributes for model parsimony without compromising the model fit. For those two priors, most of the items with $\gamma<0.4$ were detected as symmetric and most of the asymmetric items with highest asymmetry ($\gamma>0.6$) were correctly detected as asymmetric. The highly asymmetric items were also detected as asymmetric when using the Pr1 prior.

For the dataset with only symmetric items, the proportion of items correctly detected as symmetric are 0.6, 1, 1, for $J=1000$, 0.65, 0.95, 0.975, for $J=3000$ and 0.8, 1, 1, for $J=10000$, under prior specifications Pr1, Pr2 and Pr3, respectively.

\begin{figure}[h!]
 	\centering
 	\includegraphics[width=0.95\textwidth]{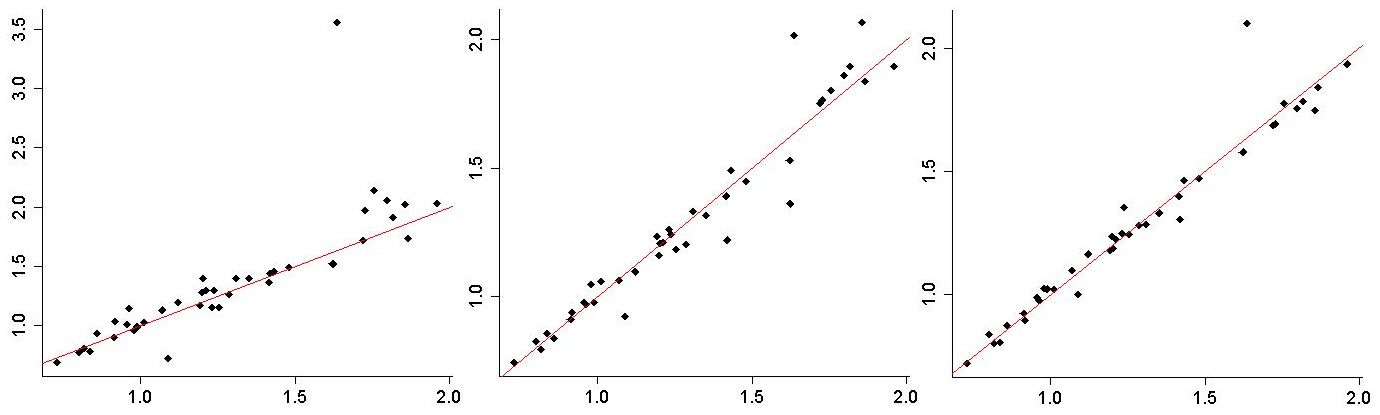}\\
    \includegraphics[width=0.95\textwidth]{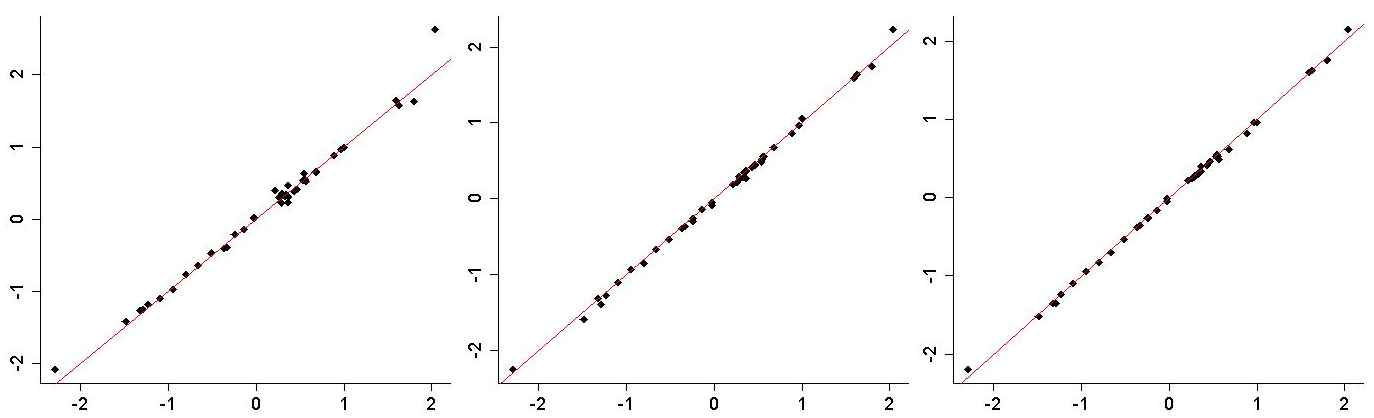}\\
  	\includegraphics[width=0.95\textwidth]{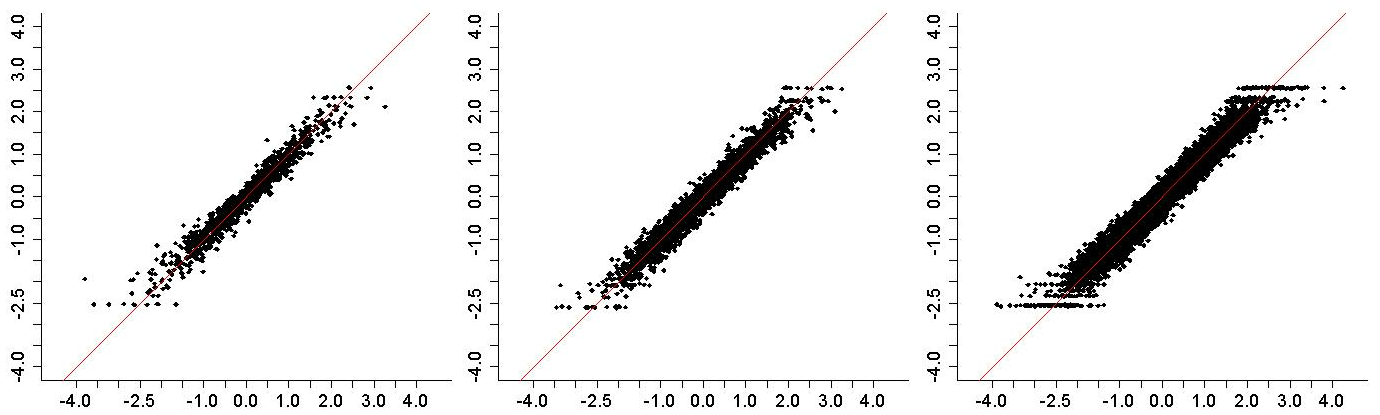}\\
  	\includegraphics[width=0.95\textwidth]{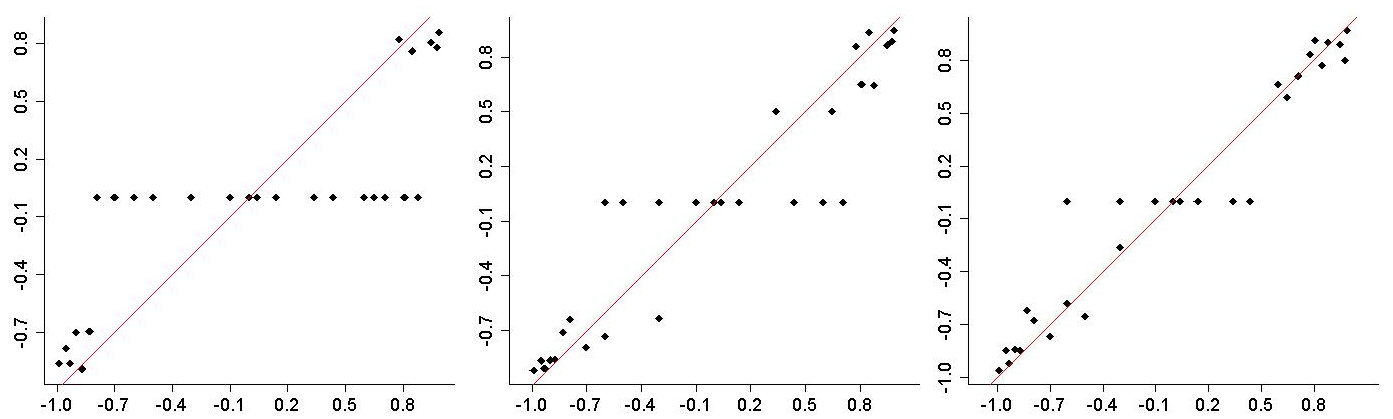}
 	\caption{True values (x-axis) versus posterior mean (y-axis) for discrimination (row 1), difficulty (row 2), ability (row 3) and skewness (row 4) parameters for sample sizes 1k (column 1), 3k (column 3) and 10k (column 3), for the first scenario, with prior Pr3.}\label{fig2}
\end{figure}

\begin{figure}[h!]
 	\centering
 	\includegraphics[width=0.95\textwidth]{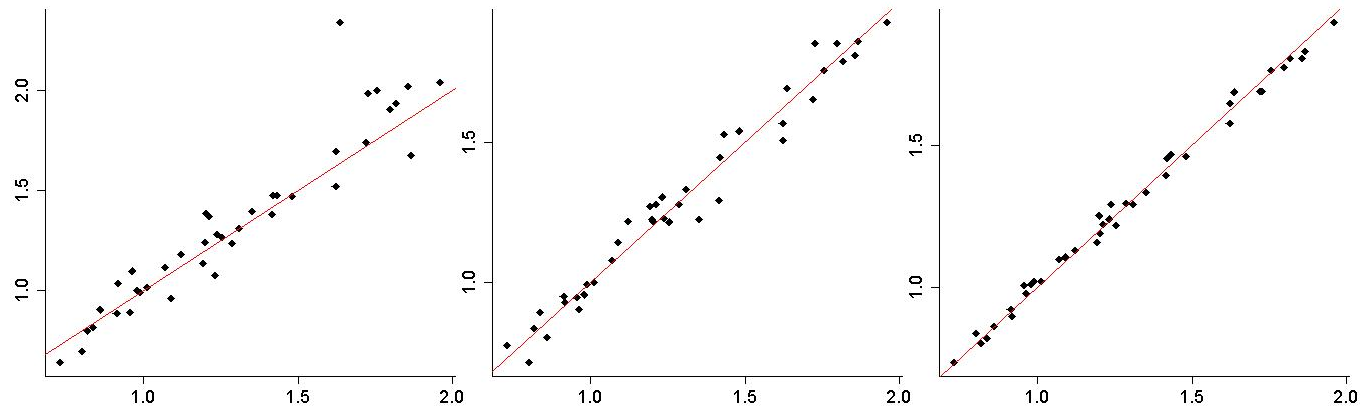}\\
    \includegraphics[width=0.95\textwidth]{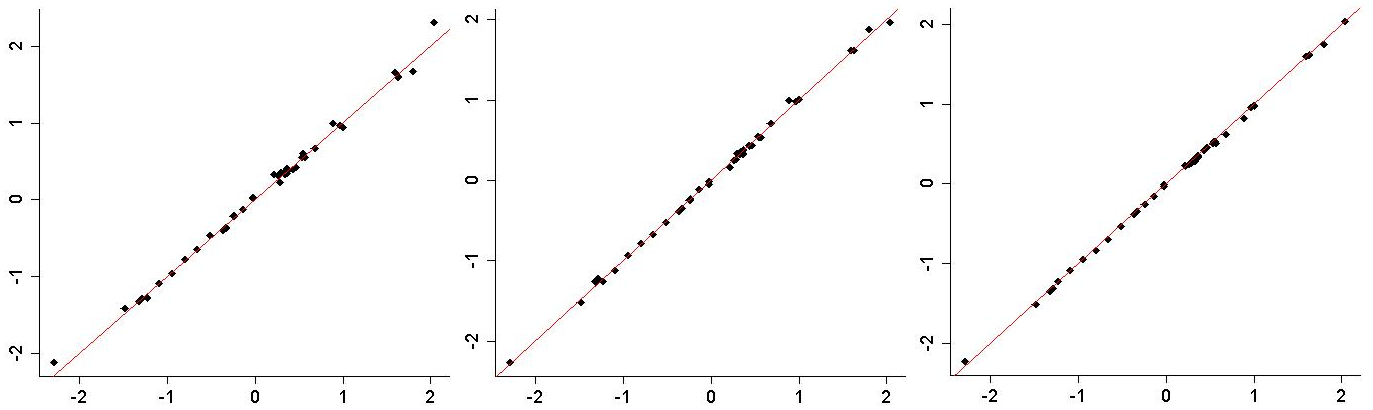}\\
  	\includegraphics[width=0.95\textwidth]{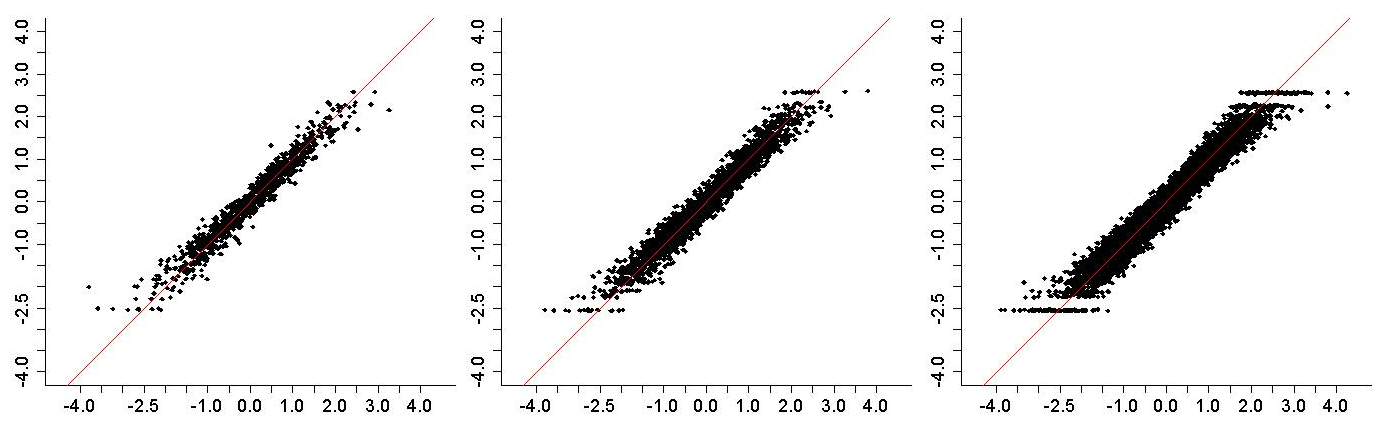}
 	\caption{True values (x-axis) versus posterior mean (y-axis) for discrimination (row 1), difficulty (row 2) and ability (row 3) parameters for sample sizes 1k (column 1), 3k (column 3) and 10k (column 3), for the second scenario, with prior Pr3.}\label{fig1}
\end{figure}

Figures \ref{fig2} to \ref{fig1} show the true values and posterior estimates for the discrimination, difficulty, abilities and skewness parameters, when a Dirichlet$(0.1, 0.01, 0.01)$ prior is assumed for the weights, for the first and second scenarios, respectively. The 2PCSNO has a very good performance in recovering the true parameters in both scenarios and for all three sample sizes. In particular, the model is efficient to correctly identify the items' skewness status even when the true model is the symmetric one.

\begin{figure}[h!]
 	\centering
 	\includegraphics[width=\textwidth]{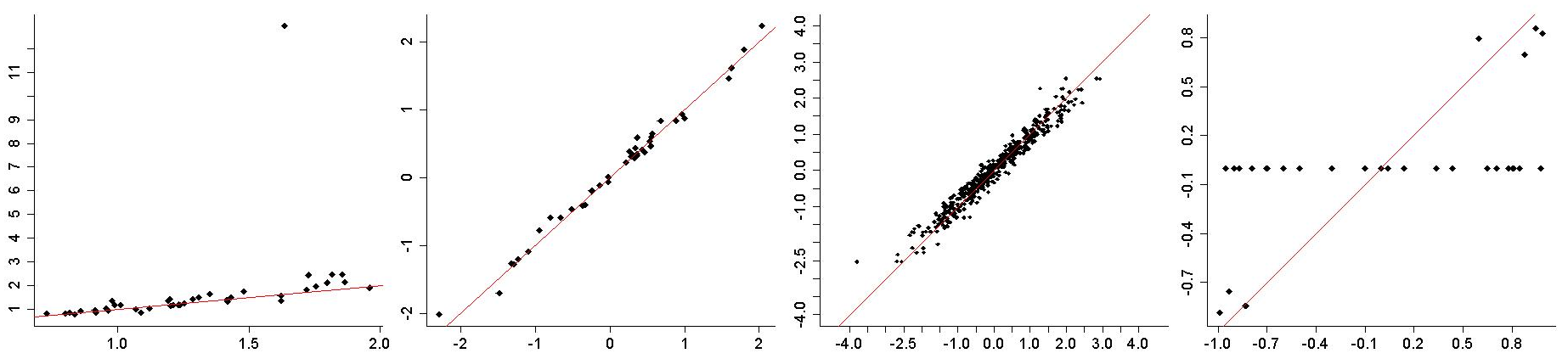}
  	\caption{True values (x-axis) versus posterior mean (y-axis) for discrimination, difficulty, ability and skewness parameters for sample size 500, for the true model with mostly asymmetric items with prior Pr3.}\label{fig3}
\end{figure}

Results show a clear impact of the sample size in the estimation of the curves' skewness. We reinforce that by showing results for $J=500$ for the dataset with mostly (27) asymmetric items. Note from Figure \ref{fig3} that only very high levels of asymmetry are detected.

\subsection{Simulation 2: Comparison to \citet{bazan2006skew}}

We shall name the model introduced by \citet{bazan2006skew} the two parameters probit skew normal ogive model (2PSNO). The two basic differences between the 2PCSNO and the 2PSNO are: $(i)$ the 2PCSNO adopts the centred parametrisation for the skew normal distribution in terms of the Pearson coefficient parameter $\gamma$ and the 2PSNO adopts the non-centred version in terms of the skewness parameter $\delta$;   $(ii)$ the 2PSNO sets a uniform prior on $[-1,1]$ for the skewness parameter $\delta$ and the 2PCSNO sets the point-mass mixture prior defined in (\ref{eq:priorGamma}) for the skewness parameter. Thus, these two approaches are two different parametrisations of the same model for the data assuming different prior specification for a set of parameters. For this reason, we are able to simulate common datasets under both models, allowing for a fair comparison.

We simulate one dataset from the model considering a test composed by 40 items of which 28 are asymmetric (being 14 positively asymmetric). We assume that the test is applied to 10,000 individuals. The same priors from Simulation 1 (with prior Pr3 for the weights) are considered here. We run the MCMC chain for 50,000 iterations with a burn-in of 1,000 for the 2PCSNO and for 100,000 iterations with a burn-in of 20,000 for the 2PSNO.

Generally speaking, the MCMC algorithm from \citet{bazan2006skew} is heavily penalised by the strong (linear or quadratic) correlation between parameters $(a_{i}^*,b_{i}^*)$ and $\delta$ (see Figure \ref{figap4}), where $a_{i}^*$ and $b_{i}^*$ are the discrimination and difficulty parameters under the non-centred parametrisation used in \citet{bazan2006skew}. As a consequence, their algorithm demands a considerably larger number of iterations than ours to achieve convergence, resulting in a considerably higher computational cost. We measure the computational cost in terms of the effective sample size (ESS) and average time per 1 effective sample (T$/$ESS, where T is the total running time of the MCMC) of the item parameters. 
The ESSs are computed with the R package CODA \citep{CODA} and considering the MCMC samples after the chosen burn-in. The MCMC for the 2PCSNO model takes 15.78 hours to run 50,000 iterations whilst the one from the 2PSNO takes 25.11 hours to run 100,000 iterations. We adopt burn-in values of 1,000 and 20,000, respectively, leading to a final sample of 49,000 for the PCSNO and 80,000 for the PSNO. Results are presented in Table \ref{tab_cost}. For the PCSNO model, only the $\gamma$ parameters of the items detected (with highest probability) as asymmetric are considered.

The trace plots of the item parameters present good convergence behaviors for the 2PCSNO model, but the same does not happen for the 2PSNO. We computed the Heidelberger and Welch's convergence diagnostic \citep{Heidel1983}, using the R package CODA, for all the chains of the item parameters. Even with a burn-in period of 20,000 adopted for the 2PSNO model, it was still possible to reject the null hypothesis of convergence for 32 parameters (1 discrimination, 11 difficulty and 20 skewness), indicating that several marginal MCMC chains did not converge. Nevertheless, the remaining MCMC iterations were used for statistical inference. On the other hand, only one (discrimination) parameter failed the test for the 2PCSNO model.

\begin{table}[h]
  \centering
  \begin{tabular}{|c|l|r|c|l|r|r|}
    \hline
               \multicolumn{3}{|c|}{2PCSNO} & \multicolumn{3}{c|}{2PSNO}  & \\ \hline
              & \multicolumn{1}{c}{ESS} & \multicolumn{1}{|c|}{T$/$ESS} &  & \multicolumn{1}{c}{ESS} & \multicolumn{1}{|c|}{T$/$ESS} & \multicolumn{1}{c|}{R.E.} \\ \hline
   $a_S$      & 1790 (763, 1713, 2976)  & 31.73s & $a_{S}^*$  & 83 \; (36, 79, 179) & 1089.10s & 34.32 \\
   $a_A$      & 1395 (132, 1468, 3089)  & 40.72s & $a_{A}^*$  & 681 (18, 653, 1830) & 132.74s  & 3.25 \\
   $b_S$      & 822 \; (473, 816, 1254) & 69.11s & $b_{S}^*$  & 24 \;  (12, 22, 51) & 3766.24s & 54.49 \\
   $b_A$      & 680 \; (155, 698, 1311) & 83.54s & $b_{A}^*$  & 492 (10, 548, 1017) & 183.73s  & 2.19 \\
   $\gamma_S$ & 2413 (2413, 2413, 2413) & 23.54s & $\delta_S$ & 23 \; (13, 20, 50)  & 3930.26s & 166.96 \\
   $\gamma_A$ & 2100 (414, 1919, 4464)  & 27.05s & $\delta_A$ & 674 (11, 686, 1720) & 134.11s  & 4.95 \\
    \hline
  \end{tabular}
  \caption{Effective sample size average (min, median, max) among items and average time (in seconds) per one effective sample of the item parameters for 2PCSNO and 2PSNO models. Subscripts $S$ and $A$ correspond to the true skewness status of the items, symmetric or asymmetric, respectively. The relative efficiency (R.E.) is the ratio of the T$/$ESS values for the 2PSNO and the 2PCSNO models.}
  \label{tab_cost}
\end{table}

Results show that the difference between the ESS averages of the two models is much larger for the symmetric items, for all the item parameters. Furthermore, whilst for the 2PCSNO the average ESS is higher for the symmetric items, for the 2PSNO it is higher for the asymmetric ones. This clearly shows the effect of the point-mass mixture prior adopted for the skewness parameter in the 2PCSNO model. For the symmetric items correctly identified as such, the respective ICC estimated by the 2PCSNO model is a mixture of ICCs with the symmetric (therefore simpler) ICC having the highest weight. The simpler structure of the symmetric ICC in turn leads to better convergence and mixing properties of the respective item parameters.

Statistical inference for some of the parameters in the 2PSNO model is unreliable given the low values of the ESS shown in Table \ref{tab_cost}. Based on those values, in order to obtain an average ESS of around 400 for parameters like $b_{S}^*$ and $\delta_S$, the MCMC chain would have to run for around 1.5M iterations, which is computationally impractical.

In the studies presented in \citet{bazan2006skew} the authors obtain results with considerably better convergence and higher ESSs than the ones presented here for the 2PSNO model. Nevertheless, whilst our simulated dataset emulates a complex situation (40 items, 10,000 individuals, 28 asymmetric items), they consider a dataset with only 6 items and 39 individuals. Although higher ESS values are reported, it is not entirely clear how the ESS values are computed and if the considered ESS represents the effective sample size. It is also not clear how these ESS values could be obtained given that their dataset consisted only of 6 items and 39 individuals, showing a lack of data information about the skewness and discrimination parameters, which usually leads to estimation problems.

\section{Application}\label{Se5}

We apply the proposed methodology to a data set from the Brazilian High School National Exam (Enem). This exam is annually applied to high school students and is organized by the Instituto Nacional de Estudos e Pesquisas Educacionais An\'{i}sio Teixeira (INEP) from the Ministry of Education (MEC). It aims to assess students who are concluding or have already concluded high school in the previous years.
Enem is composed of five sub exams: Humanities (H), Natural science (NS), Languages (Lang) and Maths (MT), each with 45 dichotomous unidimensional items, and an Essay. We consider data from the Languages exam of 2017 with a random sample of 20,000 students, excluding the 5 items related to foreign language. The complete data set is available at \emph{https://www.gov.br/inep/pt-br/acesso-a-informacao/dados-abertos/microdados/enem}.

As a test at higher grade levels, Enem is used in many universities admission process in Brazil. Thus it is important that it allows us to evaluate if students have individual skills in some specific subject area, but also if they developed the ability to use all the skills together to solve more complex tasks. In order to achieve this goal, it should include items that are able to discriminate students with high ability. The asymmetry of an item is an indicator of the item complexity and negatively asymmetric items are more efficient to discriminate students with higher abilities. We consider the proposed methodology to evaluate if the 2017 Enem exam included a significant proportion of high complex items allowing for a fair selection procedure.

The official analysis of the Enem exams is performed via the 3PL model, assuming a normal distribution with mean 500 and standard deviation 100 for the students abilities. We fit both the classical 3PNO and the 3PCSNO models, assuming a standard normal distribution for the abilities.

\begin{figure}[h!]
 	\centering
 	\includegraphics[width=\textwidth]{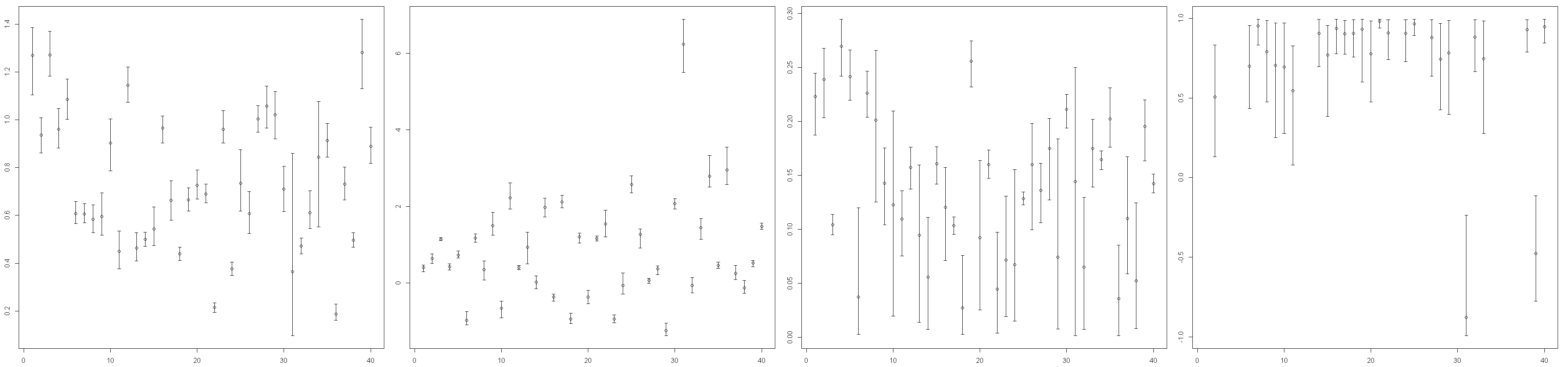}
  	\caption{Point and interval estimates (posterior mean and 95\% C.I.) of the item parameters in the Enem application. From left to right: discrimination, difficulty, guessing and skewness.}\label{fig4}
\end{figure}

\begin{figure}[h!]
 	\centering
 	\includegraphics[width=\textwidth]{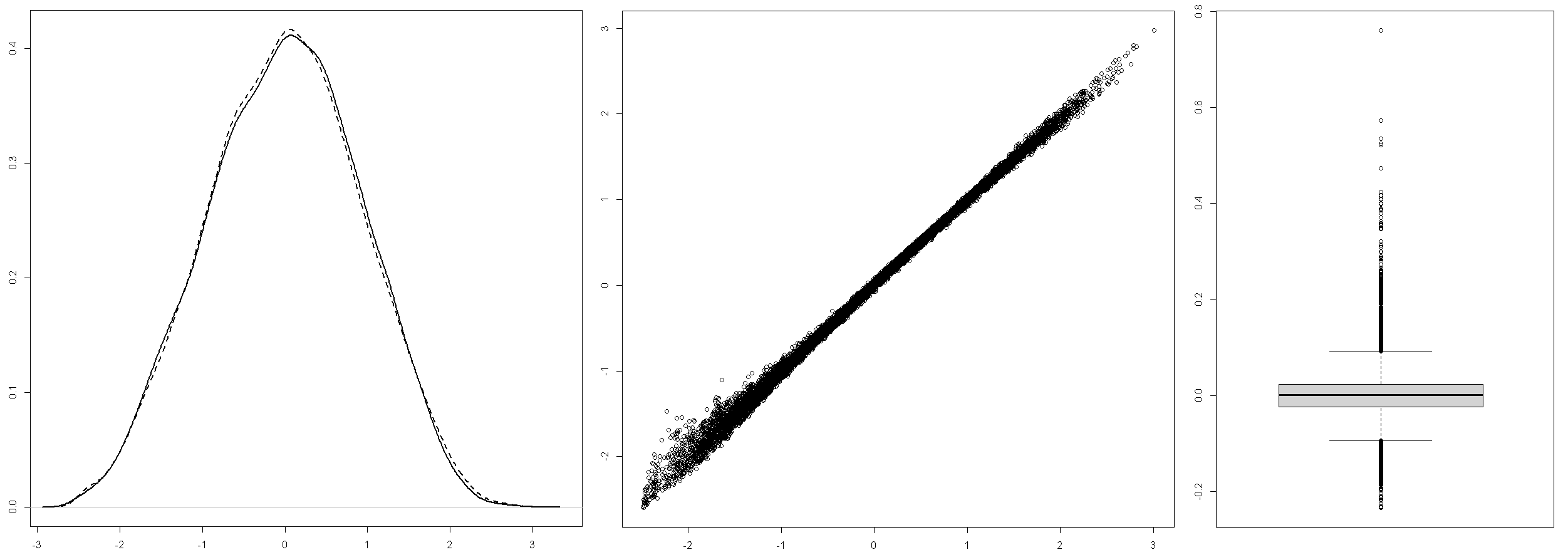}
  	\caption{Enem results. Left: distribution of the estimated abilities under the symmetric (dashed) and asymmetric (solid) models for the Enem application. Middle: estimated abilities for the symmetric (y-axis) versus asymmetric (x-axis) models. Right: box plot of the differences between the estimated abilities under the two fitted models. }\label{fig5}
\end{figure}

\begin{figure}[!h]
	\centering
	\includegraphics[width=0.8\textwidth]{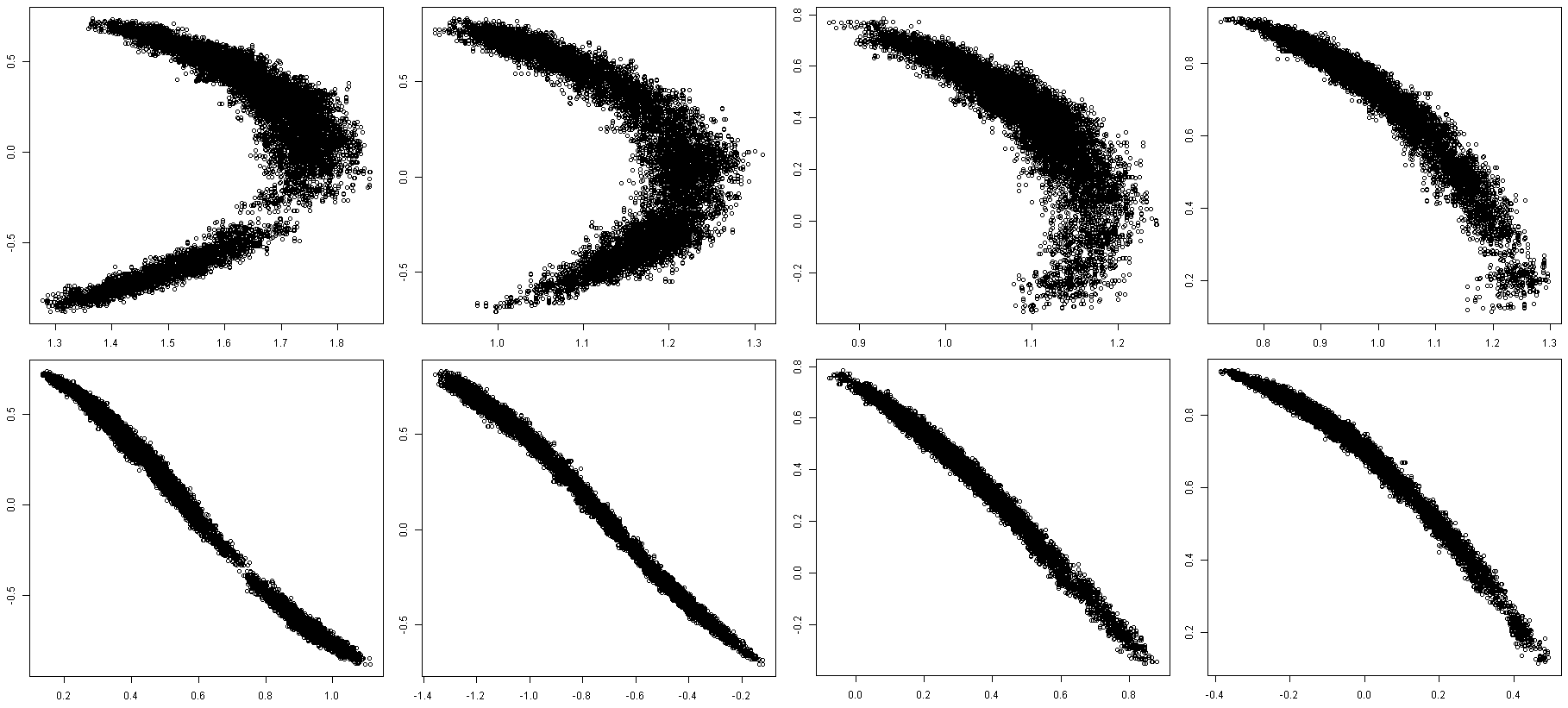}
	\caption{MCMC correlation plot between $\delta$ (y-axis) and the discrimination (top) and difficulty (bottom) parameters of items 12, 17, 28 and 29 in one of the replications for the 2PSNO model.}
	\label{figap4}
\end{figure}

We adopt the same prior distributions as in the simulated examples in Section \ref{Se4}. Results are presented in Figures \ref{fig4} and \ref{fig5}. As it was mentioned above, positively asymmetric items would help on the discrimination among students with higher ability, nevertheless, the majority of items (22) is detected to be positively asymmetric. The 3PCSNO model detected 13 symmetric items (with highest posterior probability of being symmetric), 2 negatively asymmetric items and 25 positively asymmetric items (22 with probability higher than 0.95), 14 of which with a skewness parameter estimate higher than 0.88.

The high number of detected asymmetric items suggest that the 3PCSNO is more suitable for the analysis than the tradition 2PNO. Indeed, a model comparison via DIC clearly chooses the 3PCSNO. The respective values of the DIC for the 2PCSNO and 2PNO are 886,090 and 918,760. Furthermore, a clear impact on the estimation of the abilities can be seen in the second and third plots in Figure \ref{fig5}, specially for the lowest ones. The DIC is computed by integrating out all the $D_{ij}$'s and $Z_{i}$'s latent variables and treating the $\theta_j$'s as parameters, more specifically:
\begin{eqnarray*}\label{eq:dic}
\ds DIC &=& -4E_{\Psi|\mathbf{y}}\left[ l(\Psi, \mathbf{y})\right] +2l(\hat{\Psi}, \mathbf{y}); \\
l(\Psi, \mathbf{y}) &=& \sum_{i=1}^I\sum_{j=1}^J \log\left(p_{ij}(m_{ij},\gamma_i)^{y_{ij}}(1-p_{ij}(m_{ij},\gamma_i))^{1-y_{ij}}\right); \\
p_{ij}(m_{ij},\gamma_i) &=& c_i+(1-c_i)\Phi_{CSN}(a_i(\theta_j-b_i),\gamma_i).
\end{eqnarray*}
The expression for $l(\hat{\Psi}, \mathbf{y})$ is obtained by replacing $\theta_j$, $a_i$, $b_i$, $c_i$ and $\gamma_i$, $\forall$ $(i,j)$, by their respective posterior means, in the expression of $l(\Psi, \mathbf{y})$.

\section{Final remarks}
\label{Se6}

This paper proposed a flexible Bayesian IRT model for dichotomous items to account for asymmetric item characteristic curves. The main contributions of the proposed methodology lies on the fact that the model is flexible yet parsimonious due to the point-mass mixture prior adopted for the skewness parameter. This allows for an analysis under both model selection and model averaging approaches. From the practical point of view, it provides an efficient way to validate items by allowing them to be more easily classified as symmetric, negatively or positively asymmetric thus defining its complexity level.

A particular parametrisation of the model has shown to be appealing in terms of both parameter interpretation and computational efficiency. Furthermore, an efficient MCMC algorithm was proposed to sample from the posterior distribution of all the unknown quantities in the model.

A sensitivity analysis for the prior on the weights of the mixture was performed through simulated examples, which also highlighted the efficiency of the MCMC algorithm. Finally, the applicability of the proposed methodology was illustrated in the analysis of data coming from a large scale educational assessment exam in Brazil. The detection of a majority of negatively asymmetric items was particularly interesting.

Finally, the proposed methodology can be extended to consider skew models for ordinal response data. For example, one may replace the normal cdf in the graded Response Model (GRM) of \citet{samejima1969estimation} by the cdf of the centred skew normal. In this case, a straightforward extension would consider a common skewness parameter for all the cfd's of the same item. More general forms would have to be carefully designed to preserve the monotonic relation of all those cdf's. This is a possible direction for future works.

\bibliographystyle{apa}
\bibliography{bibliografia}

\newpage

\section*{Tables and Figures}

\newpage

\textcolor[rgb]{1.00,1.00,1.00}{.}

\newpage

\textcolor[rgb]{1.00,1.00,1.00}{.}

\newpage

\section*{Appendix A}\label{Appendix}

\newpage

\section*{Appendix B}

The proposed MCMC algorithm is as follows:

\textbf{Step 1}: Simulate $(Z_i,w_i)$, for $i = 1, \ldots, I$:

\begin{itemize}
	\item Draw $w_i^{(*)}$ from $w_i \sim Dirichlet(\alpha_0,\alpha_1,\alpha_2)$.
	\item Draw $Z_i^{(*)}$ from $Z_i|w_i^{(t)}$ $\sim Multinomial(1,w_i^{(t)})$, where $w_i = (w_{i0}, w_{i1},w_{i2})$.
	\item Accept $(Z_i^{(t)}, w_i^{(t)})$ = $(Z_i^{(*)}, w_i^{(*)})$ with probability $\min\{1, R_{(Z,w)_i}\}$, where
\end{itemize}
\begin{equation*}
R_{(Z,w)_i} = \frac{\prod^{L_j}_{j=1}\Phi_{CSN}(m_{ij}^{(t)},\gamma_i^{(*)})^{y_{ij}}[1 - \Phi_{CSN}(m_{ij}^{(t)},\gamma_i^{(*)})]^{1 - y_{ij}}}{\prod^{L_j}_{j=1}\Phi_{CSN}(m_{ij}^{(t)},\gamma_i^{(t)})^{y_{ij}}[1 - \Phi_{CSN}(m_{ij}^{(t)},\gamma_i^{(t)})]^{1 - y_{ij}}},
\end{equation*}
where $\gamma_i^{(*)} = 0{Z_{i0}^{(*)}} + \gamma_{i-}{Z_{i1}^{(*)}}+ \gamma_{i+}{Z_{i2}^{(*)}}$ and $\gamma_i^{(t)} = 0{Z_{i0}^{(t)}} + \gamma_{i-}{Z_{i1}^{(t)}}+ \gamma_{i+}{Z_{i2}^{(t)}}$, and where $L_j$ refers to the items for which $D_{ij}$ = 0.

\textbf{Step 2}: Simulate $\gamma_{i-}$ and $\gamma_{i+}$, for $i = 1, \ldots, I$:

\begin{itemize}
	\item \textit{If} $Z_{i0} = 1$, then make $\gamma_{i-}^{(t)} = \gamma_{i-}^{(t-1)}$ and $\gamma_{i+}^{(t)} = \gamma_{i+}^{(t-1)}$;
	
	\item \textit{If} $Z_{i1} = 1$, then draw $\gamma_{i-}^{(*)}$ from $N(\gamma_{i-}^{(t-1)}; \tau_{\gamma_{i-}})$, and make $\gamma_{i+}^{(t)} = \gamma_{i+}^{(t-1)}$;
	
	\begin{itemize}
		\item Accept $\gamma_{i-}^{(t)} = \gamma_{i-}^{(*)} $ with probability $\min\{1, R_{\gamma_{i-}}\}$, where
	\end{itemize}	
	\begin{equation*}
	R_{\gamma_{i-}} = \frac{\left[\prod^{L_j}_{j=1}\Phi_{CSN}(m_{ij}^{(t)},\gamma_{i-}^{(*)})^{y_{ij}}[1 - \Phi_{CSN}(m_{ij}^{(t)},\gamma_{i-}^{(*)})]^{1 - y_{ij}}\right] f_B(-\gamma_{i-}^{(*)};\alpha,\beta)}{\left[\prod^{L_j}_{j=1}\Phi_{CSN}(m_{ij}^{(t)},\gamma_{i-}^{(t)})^{y_{ij}}[1 - \Phi_{CSN}(m_{ij}^{(t)},\gamma_{i-}^{(t)})]^{1 - y_{ij}}\right]f_B(-\gamma_{i-}^{(t)};\alpha,\beta)},
	\end{equation*}
	where $L_j$ refers to the items for which $D_{ij}$ = 0.
	\item \textit{If} $Z_{i2} = 1$, then draw $\gamma_{i+}^{(t)}$ from $N(\gamma_{i+}^{(t-1)}; \tau_{\gamma_{i+}})$, and make $\gamma_{i-}^{(t)} = \gamma_{i-}^{(t-1)}$;
	
	\begin{itemize}
		\item Accept $\gamma_{i+}^{(t)} = \gamma_{i+}^{(*)} $ with probability $\min\{1, R_{\gamma_{i+}}\}$, where
	\end{itemize}	
	\begin{equation*}
	R_{\gamma_{i+}} = \frac{\left[\prod^{L_j}_{j=1}\Phi_{CSN}(m_{ij}^{(t)},\gamma_{i+}^{(*)})^{y_{ij}}[1 - \Phi_{CSN}(m_{ij}^{(t)},\gamma_{i+}^{(*)})]^{1 - y_{ij}}\right] f_B(\gamma_{i+}^{(*)};\alpha,\beta)}{\left[\prod^{L_j}_{j=1}\Phi_{CSN}(m_{ij}^{(t)},\gamma_{i+}^{(t)})^{y_{ij}}[1 - \Phi_{CSN}(m_{ij}^{(t)},\gamma_{i+}^{(t)})]^{1 - y_{ij}}\right]f_B(\gamma_{i+}^{(t)};\alpha,\beta)},
	\end{equation*}
\end{itemize}
where $L_j$ refers to the items for which $D_{ij}$ = 0.

\textbf{Step 3}: Simulate $(a_i, b_i)$, for $i = 1, \ldots, I$:

\begin{itemize}
	\item Draw $(a_i,b_i)^{(*)}$ from a bivariate normal distribution $N_2((a_i \quad b_i)^{'(t-1)};\Sigma_{ab})$,
	\item Accept the pair $(a_i,b_i)^{(t)}$ = $(a_i,b_i)^{(*)}$ with probability $\min\{1, R_{(a,b)_i}\}$, where
\end{itemize}
\begin{eqnarray*}
	R_{(a,b)_i} &=&\frac{\prod^{L_j}_{j=1}\Phi_{CSN}(a_i^{(*)}(\theta_i^{(t)}-b_i^{(*)}),\gamma_i^{(t)})^{y_{ij}}[1 - \Phi_{CSN}(a_i^{(*)}(\theta_i^{(t)}-b_i^{(*)})),\gamma_i^{(t)})]^{1 - y_{ij}}}
	{\prod^{L_j}_{j=1}\Phi_{CSN}(a_i^{(t-1)}(\theta_i^{(t)}-b_i^{(t-1)}),\gamma_i^{(t)})^{y_{ij}}[1 - \Phi_{CSN}(a_i^{(t-1)}(\theta_i^{(t)}-b_i^{(t-1)}),\gamma_i^{(t)})]^{1 - y_{ij}}}\\ [0.5cm]
	&  & \;\;\;\;\;\;\;\;\;\;\;\;\; \times\;\frac{f_{TN}(a_i^{(*)};\tau_a)f_N(b_i^{(*)};\tau_b)} {f_{TN}(a_i^{(t-1)};\tau_a)f_N(b_i^{(t-1)};\tau_b)},
\end{eqnarray*}
where  $f_N(;\tau_b)$ and $f_{TN}(;\tau_a)$ denote, respectively,  the p.d.f. of a normal distribution and a truncated normal distribution on $\mathds{R}^+$, $\tau_a$ and $\tau_b$ are tuning parameters, and $L_j$ refers to the items for which $D_{ij}$ = 0.
\\
\textbf{Step 4}: Simulate $\theta_j$, for $j = 1, \ldots, J$:

\begin{itemize}
	\item Draw $\theta_j^{(*)}$ from a normal distribution $f_N(\theta_j^{(t-1)};\sigma_{\theta})$.
	\item Accept $\theta_j^{(t)}$ = $\theta_j^{(*)}$ with probability $\min\{1, R_{\theta_j}\}$, where
\end{itemize}
\begin{equation*}
R_{\theta_j} = \frac{\left[\prod^{L_i}_{i=1}\Phi_{CSN}(a_i^{(t)}(\theta_i^{(*)}-b_i^{(t)}),\gamma_i^{(t)})^{y_{ij}}[1 - \Phi_{CSN}(a_i^{(t)}(\theta_i^{(*)}-b_i^{(t)})),\gamma_i^{(t)}]^{1 - y_{ij}}\right]f_N(\theta_j^{(*)};\tau_{\theta})}
{\left[\prod^{L_i}_{i=1}\Phi_{CSN}(a_i^{(t)}(\theta_i^{(t-1)}-b_i^{(t)}),\gamma_i^{(t)})^{y_{ij}}[1 - \Phi_{CSN}(a_i^{(t)}(\theta_i^{(t-1)}-b_i^{(t)}),\gamma_i^{(t)})]^{1 - y_{ij}}\right]f_N(\theta^{(t-1)};\tau_{\theta})},
\end{equation*}
where $L_i$ refers to the items for which $D_{ij}$ = 0.

\textbf{Step 5}: Simulate $D_{ij}$, for $i = 1, \ldots, I$:

\begin{itemize}
	\item If $Y_{ij}=0$, $P(D_{ij}=0|Y_{ij}=0,\cdot)=1$;
    \item If, $Y_{ij}=1$, compute $r_{ij}=\frac{c_i}{c_i + (1-c_i)\Phi_{CSN}(m_{ij}) )}$ and simulate $(D_{ij}|\cdot)\sim Ber(r_{ij})$.
\end{itemize}

\textbf{Step 6}: Simulate $c_i$, for $i = 1, \ldots, I$:

\begin{itemize}
	\item Draw $c_i \sim Beta\bigg(\sum_{j=1}^JZ_{ij} + \alpha_c, J - \sum_{j=1}^JZ_{ij} + \beta_c\bigg)$,
\end{itemize}
where $\alpha$ and $\beta$ are prior hyperparameters.

\end{document}